# A frequency shift compensation method for light shift and vapor-cell temperature shift in atomic clocks

Dou Li[1,2,*], Kangqi Liu[1,2,*], Linzhen Zhao[1,2], Songbai Kang[1, †]

[1] *Key Laboratory of Atomic Frequency Standards,*
*Innovation Academy for Precision Measurement Science and Technology, Chinese Academy of Sciences,*
*Wuhan 430071, China*

[2] *University of Chinese Academy of Sciences, Beijing 100049, China*

Light shift and vapor-cell temperature shift are the two most significant factors dominating the long-term instability of compact atomic clocks. Due to the different physical mechanisms, there is not yet a solution that can effectively suppress the frequency shifts induced by these two effects. Here, we propose a 'resonance-offset' locking approach that compensates for the two physical frequency shifts. In this approach, the additional offset locking shift can effectively counteract the atomic resonance shifts arising from changes in vapor-cell temperature and light power, reducing the net impact on the clock's frequency to nearly zero. We have demonstrated this strategy on the 778 nm Rubidium two-photon optical frequency standard, successfully compensating for light shift and cell-temperature shift, respectively. This general method is particularly appealing for compact vapor-cell microwave and optical atomic clocks designed for the excellent stability rather than accuracy.

## I. INTRODUCTION

Compact atomic clock technologies based on atomic vapor cells, such as rubidium atomic clocks [1-2] and chip-scale coherent population trapping (CPT) atomic clocks [3-4], have gained widespread applications in fields like navigation [5], time network synchronization [6], and power grid management due to their small size, weight, and power consumption [7]. Moreover, higher-performance vapor-cell atomic clock technologies, such as laser-pumped rubidium clocks (including continuous-wave [8] and pulsed-operation [9-10]), high-performance CPT clocks [11], and vapor-cell optical clocks [12-14], continue to be hot topics in next-generation atomic clock research. However, those vapor-cell atomic clocks' long-term stability are generally dominated by two environmental factors: the interrogation light power [12,15-18,] and the temperature of the vapor cell [19-22]. Researchers have always sought to suppress these frequency shifts through different technical approaches to enhance the long-term stability of atomic clocks. A primary method involves minimizing changes in the environmental parameters ($\Delta X$) of the atomic clocks. However, directly suppressing $\Delta X$ often requires strict control measures. For instance, space-borne rubidium clock typically needs the vapor cell's temperature to be maintained within approximately 1 mK over one day to achieve a stability of about $1\times10^{-15}$ @1day [23]. Similarly, the two-photon Rb optical frequency standard (2hv-ROFS) requires the relative stability of the interrogation light power to be maintained at 1 ppm to ensure a daily stability index at the $1\times10^{-15}$ level [12]. Besides, methods to suppress $\Delta X$ can significantly

---

[+]These two coauthors contributed equally.

[†]kangsongbai@apm.ac.cn

increase the size, weight, and power consumption (SWaP), which are critical concerns for compact atomic clock technology.

Mitigating the frequency shift coefficient or sensitivity represents another effective strategy. A key idea is 'frequency compensation' which aims to counterbalance frequency shifts due to $\Delta X$ with an additional opposive shift. For rubidium atomic clocks, a mixed-buffer-gas approach reduces the vapor-cell temperature-dependent frequency shift coefficient by employing two buffer gases (e.g., $N_2$ and $CH_4$), each of which creats opposite temperature shifts relative to the Rb transition [20,24-25]. For 2hv-ROFS, a two-color light approach has been developed to mitigate light shift by using two different light colors with opposing shift polarities [26]. These opposing shifts cancel each other out, reducing the total light-induced shift. Additionally, light shift compensation through an asymmetry-generated frequency shift in the atomic resonance signal has been explored for the CPT clocks by varying the polarization or detuning of the two-color resonance light to offset frequency shifts caused by light intensity [27-30]. Recently, the combined error signal method, which involves power modulation or dual interrogation, has been extensively researched for compensating for power-induced light shifts in vapor-cell atomic clocks under continuous-wave or Ramsey schemes [31-34]. Nevertheless, all of the shift-suppressing techniques primarily work for either light shift or vapor-cell temperature shift, not a general solution for both.

Here, we introduce a simple method for compensating the two different shifts using a 'resonance-offset locking (ROL)' technique. This approach involves a locking point offset from the atomic resonance peak which introduces an offset-locking shift. This additional locking shift is dependent on the amplitude of the atomic resonance and can be used for counteracting the physical atomic shift induced by changes in light power or cell temperature. As this technique is an easy operation for the post-atomic resonance, it is particularly suitable for compact atomic clocks without complicating the atomic clock system.

The structure of this paper is as follows: Section II outlines the ROL method's compensation principle. Section III introduces the setup of the 778 nm 2hv-ROFS platform for the ROL method's shift mitigation. Experimental outcomes for both light and vapor-cell temperature shift compensation on this platform are discussed in Section IV. Lastly, Section V offers a discussion and concludes our findings.

## II. PRINCIPLE

In Figure 1, $S_i$ denotes the error signals of atomic transitions under $X_i$ (either laser power or cell temperature) with a slope of tag($\varepsilon_i$) (i=1,2). The output frequency corresponding to on-resonance locking (i.e., when $V_{offset}$ = 0) is represented as $f_i$. The term $\Delta f_{\Delta X} = f_2 - f_1$ signifies the frequency shift resulting from $\Delta X = X_2 - X_1$. The two error signal curves intersect at a special cross point ($f_{offset}$, $V_{offset}$), where the clock's output frequency $f_{offset}$ does not change with $\Delta X$. Thus, we consider $V_{offset}$ as the optimized ROL point where perfect shift compensation can be achieved.

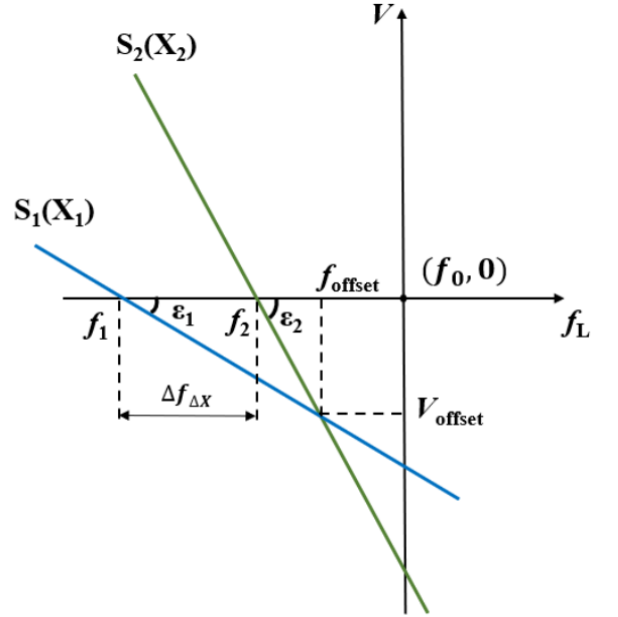

FIG. 1. Illustration of error signal's linear section. Error signals $S_1(X_1)$ (blue) and $S_2(X_2)$ (green) operated under the environmental parameters $X_1$ and $X_2$, with the slopes of $\tan(\varepsilon_1)$ and $\tan(\varepsilon_2)$ respectively. The cross point ($f_{\text{offset}}$, $V_{\text{offset}}$) is the optimal ROL point that the clock's output is insentive to $\Delta X = X_2 - X_1$.

Let's assume that the slope tag($\varepsilon_i$) of the atomic error signal is related to the parameter $X_i$ as follows:

$$tag(\varepsilon_i) = \kappa X_i^r \qquad (1)$$

where is $\kappa$ and r (r $\neq$ 0) can be arbitrary. The clock's shift induced by $\Delta X$ should be proportional to this relationship, which is easily satisfied when $\Delta X$ is much smaller than $X_1$ and $X_2$. So we get

$$\Delta f_{\Delta x} = f_2 - f_1 = \alpha(X_2 - X_1) = \alpha \Delta X \qquad (2)$$

where $\alpha$ is defined as the parameter $X$'s shift coefficient. We note that equations (1) and (2) are the prerequisites of the offset-locking compensation.
According to Fig. 1, we have

$$f_{offset} - f_1 = \frac{V_{offset}}{tag(\varepsilon_1)|_{X=X_1}} \qquad (3)$$

$$f_{offset} - f_2 = \frac{V_{offset}}{tag(\varepsilon_2)|_{X=X_2}}. \qquad (4)$$

From the equations (2), (3) and (4), we derive

$$V_{offset} = \frac{-\alpha \Delta X}{\frac{1}{tag(\varepsilon_2)|_{X=X_2}} - \frac{1}{tag(\varepsilon_1)|_{X=X_1}}}$$
$$= \frac{-\alpha \Delta X}{\left(\frac{1}{tag(\varepsilon_i)}\right)' |_{X=X_1} \Delta X}. \qquad (5)$$

Therefore,

$$V_{offset} = -\frac{\kappa \alpha}{r} X^{r+1} \qquad (6)$$

## III. EXPERIMENTAL SETUP

In this work, we employ the ROL technique to compensate for the light shift and vapor-cell temperature shift in the 778 nm 2hv-ROFS, respectively. The approximately 2 nm far-detuning from the 778 nm two-photon Rb 5S-5D transition ensures that its linewidth remains unaffected by power broadening due to changes in probe light power and vapor-cell temperature [35]. This unique property enables us to perform experiments compensating for the light shift or cell temperature shift independently, without interference between the two effects. Accordingly, we used the amplitude of the atomic signal as an indicator of the slope of the atomic error signal for subsequent experiments. It is also noteworthy that our focus was on assessing $\Delta X$ induced frequency shift sensitivity rather than precisely measuring the absolute frequency of the systematic effect.

Figure 2a depicts the setup for the 778 nm 2hv-ROFS. A 1556 nm narrow-linewidth laser (~100 Hz instantaneous linewidth) is split into two beams via a fiber beam splitter. One beam is directed through an Erbium-Doped Fiber Amplifier (EDFA) and a Second Harmonic Generator (SHG) to generate approximately 50 mW of 778 nm light, and the other beam beats with an optical frequency comb referenced to either a H-maser or a ultra-stable laser stabilized on a super-cavity. The 778 nm beam passes through a fiber-type Variable Optical Attenuator (VOA, ~1 kHz bandwidth) for intentional adjustment in light-shift suppression experiments. A Glan-Taylor prism (GTP) and a free-space 9:1 Beam Splitter (BS) ensure the probe beam's high linear polarization for interrogation and monitoring of input light power. The interrogation beam, approximately 800 μm in diameter and with ~ 15 mW of power, is directed into a cubic $^{87}$Rb glass vapor cell. This cell is heated to around 100 °C to maintain a high density of Rb vapor. The incoming probe light is precisely counter-reflected by a 778 nm high-reflective coating on the cell's rear, which is essential for observing the two-photon spectrum. A Photomultiplier Tube (PMT) captures the 420 nm fluorescence through a pair of aspherical lenses, with two layers of magnetic shielding isolating environmental magnetic fields and achieving a shielding factor of about 1500. Figure 2b illustrates the typical 420 nm atomic fluorescence signal, with a fitting full width at half maximum (FWHM) of approximately 618 kHz (778 nm carrier).

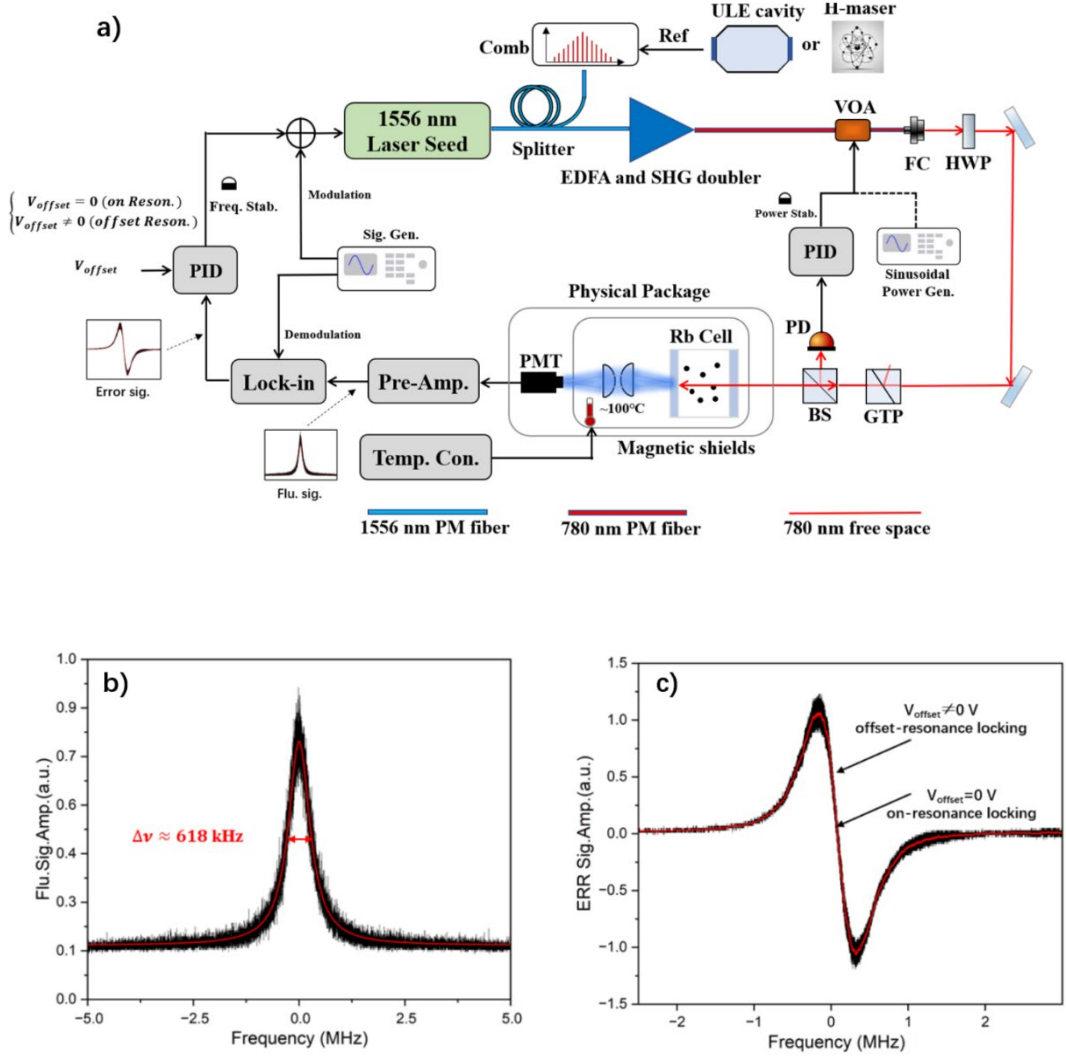

FIG. 2. (a). Schematic of the 2hν-ROFS platform. Terms used: VOA, variable optical attenuator; FC, fiber collimator; HWP, half-wave plate; PD, photodiode; PID, proportional-integral-derivative; BS, beam splitter; GTP, Glan–Taylor prism; PMT, photomultiplier tube; Lock-in, lock-in amplifier; Pre-Amp, preamplifier; EDFA, erbium-doped fiber amplifier; SHG, second harmonic generation; FG, function generator; and Temp. Con. Temperature controller. (b).420 nm atomic fluorescence signal and a Lorentz-function fitting (red) with a linewidth of 618 kHz; (c).corresponding error signal monitor by output of Lock-in and 100-point smoothing line (red).

To obtain the atomic discriminator or error signal, we modulate the source laser at approximately 10 kHz with a frequency deviation of about 50 kHz. The 420 nm fluorescence atomic signal is then amplified by a Preamplifier (PA) and synchronously demodulated by a Lock-In Amplifier. Figure 2c displays the error signal corresponding to the atomic fluorescence signal. This error signal is then fed back to the laser via a Proportional-Integral-Derivative (PID) servo for loop locking. The PID's offset setting ($V_{offset}$) allows for the manual selection between on-resonance locking ($V_{offset} = 0$) and the offset locking points ($V_{offset} \neq 0$).

## IV. EXPERIMENT RESULTS

### A. Light shift

We begin by examining the power-induced light shift characteristics of the 2hv-ROFS. In Fig. 3(a) and (b), the results showcase how changes in probe light power influence the atomic signal amplitude and light shift. In panel a, the red dashed fitting line, with an $R^2$ value of 0.99992, demonstrates a quadratic increase in signal amplitude in response to rising probe power. This is in alignment with the transition probability being proportional to the square of the laser power [35]. In panel b, the AC Stark effect causes the output frequency to linearly correlate with the input power, yielding a fractional shift coefficient of 414(10) Hz/mW @ 778 nm for a laser beam diameter of approximately 800 μm. These two features show that the laser-power-induced light shift meets the ROL compensation prerequisites.

To ascertain the optimal ROL point, we tested various $V_{offset}$ settings against the resultant light shift responses. Figure 4a reveals the system's output frequency across different $V_{offset}$ values under varying probe light powers. Since the chosen offset locking points are close to the peak on-resonance point, the output frequency changes linearly with $V_{offset}$, having a linear coefficient of approximately 100 Hz/mV (dashed line). However, the light shifts caused by power variations are relatively small and thus cannot be clearly distinguished in the graph. To more clearly demonstrate the influence of the offset locking points on the light shift coefficient, we set the starting frequency of each light shift coefficient curve (under 14.0 mW power) to "zero offset" and redraw the measurement data, as demonstrated in Figure 4b. Compared to resonance locking ($V_{offset} = 0$ mV, blue dots), frequencies taken at a $V_{offset}$ of 25.5 mV (black squares) show minimal power-dependent frequency variance, suggesting effective compensation through offset locking. Furthermore, data operated under a $V_{offset}$ of 65.5 mV (green diamonds) indicates 'overcompensation,' reversing the light shift effect, while data run with a $V_{offset}$ of -14.5 mV (yellow triangls) presents a more significant coefficient than the baseline scenario. These results pinpoint a $V_{offset}$ of 25.5 mV as ideal for stabilizing the output frequency against light power fluctuations, effectively achieving a 'zero' light frequency shift coefficient.

To further substantiate the real-time compensation effect on light shift, we introduced a controlled disturbance to the probe power and analyzed the stability performance of the optical frequency standard with different offset voltages. Figure 5a shows the manipulation of the VOA with a waveform generator to create probe light exhibiting a sawtooth wave. This 'disturbed' probe light had an average optical power of 14 mW, with a 100-second period and a valley-to-peak amplitude of 1 mW. Figures 5b and 5c display the real-time output frequency of the 2hv-ROFS and its stability throughout power-disturbed operation at different offset-locking voltages. It is important to highlight that this measurement employs the optical frequency comb referenced

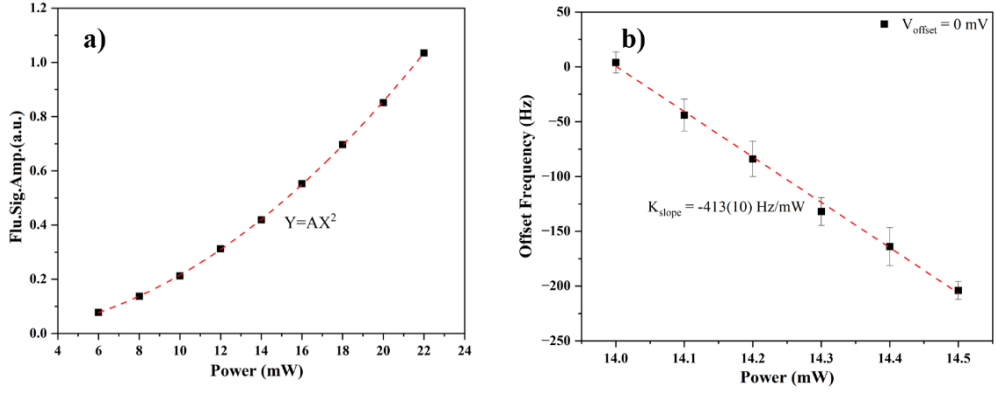

FIG. 3. (a) Experimetally measured 420 nm fluoresonce amplitude depends on the input laser power. The red dash line is the fitting result based on Y=AX$^2$, where R$^2$= 0.99992. (b) Experimentally measured light power shift (on resonance locking, V$_{offset}$ = 0 V), the fitting slope is -414 (10) Hz/ mW, where R$^2$= 0.9976.

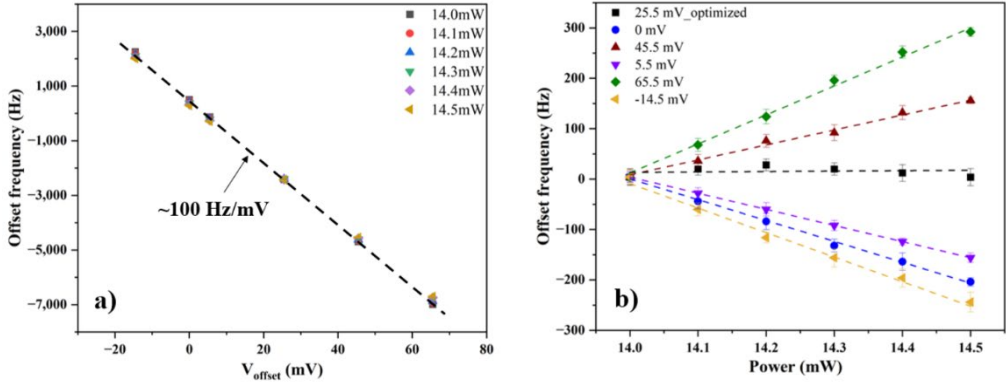

FIG. 4 (a) Measured offset frequency depends on the offset locking points V$_{offset}$, under different interrogation laser power. Frequency data obtained when V$_{offset}$ = 0 V is set to "Zero" offset frequency. (b) Measured offset frequency depends on the laser power operated with different offset locking points V$_{offset}$. Frequency data run under 14.0 mW laser power condition is set to "Zero" offset frequency.

to a super-cavity-stabilized laser, with the stability shown in Figure 5c (brown). For comparison, the undisturbed condition with a constant power of 14 mW is also provided (black). Under on-resonance locking (V$_{offset}$ = 0 mV, blue), the output frequency clearly synchronizes with the changes in light power, with its corresponding frequency stability being most severely degraded at an averaging time of 50 seconds, reaching 2E-13. In the case of overcompensation at a V$_{offset}$ of 65.5 mV, the output frequency is adversely affected, as evidenced by a stability drop to 3E-13 @ 50s. In stark contrast, 'optimized' offset locking at V$_{offset}$ = 25.5 mV effectively neutralizes the light power disturbances, achieving stability of 3E-14 @ 50s consistent with the undisturbed operation. It can conservatively be concluded that the offset locking approach suppressed light shift sensitivity by at least an order of magnitude.

**B. Vapor-Cell temperature shift**

Our empirical research has focused on the vapor-cell temperature sensitivity of the 2hv-ROFS, with the probe light power set at 15 mW. Illustrated in Figures 5(a) and (b), we investigated both the amplitude of the atomic fluorescence signal and the frequency shifts induced by cell temperature, noting a clear correlation with the vapor cell's operational temperature. In panel a, the amplitude of the fluorescence signal nearly increases linearly with the rise in cell temperature. While ideally following the Antoine equation, the small temperature variation range (104 °C to 106 °C) allows for a linear approximation. In panel b, our linearly fitted vapor-cell temperature fractional shift sensitivity is about 1070 (22) Hz/°C @ 778 nm (on-resonance locking, $V_{offset}$ = 0 mV), approximately 2 times larger than the coefficient reported in Ref. [12]. We think this discrepancy may arise from either servo errors in the locking loop or/and the residal magnetic field generated by the cell-heating current, and another possibility is from the changes in the polarization of the probing light due to the tension changes in the transparent face of the cubic vapor cell as the temperature varies. Despite these factors, the measurement accurately defines the vapor-cell temperature sensitivity and does not detract from demonstrating the efficacy of compensation. Consequently, the identified cell temperature shifts meet the requirements for implementing ROL compensation.

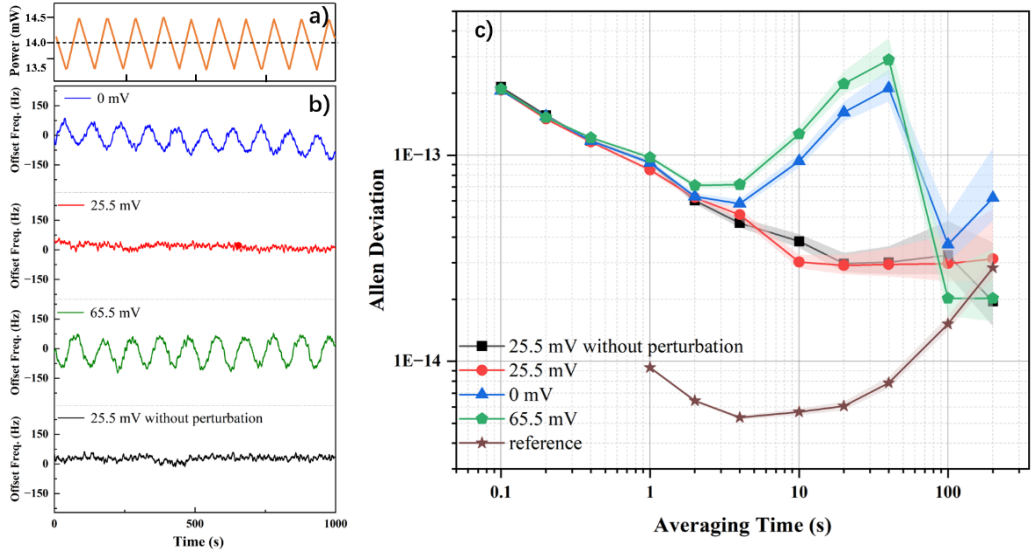

FIG. 5. (a) "disturbed" probe power's real-time variation. (b) 100-point averaged output freuqency operated under on-resonance locking (blue), optimal offset locking (red), overcompensating locking (green), and no power disturbance condition (black). (c) the corresponding stability performances for the four conditions in (b), brown trace is the optical reference stability.

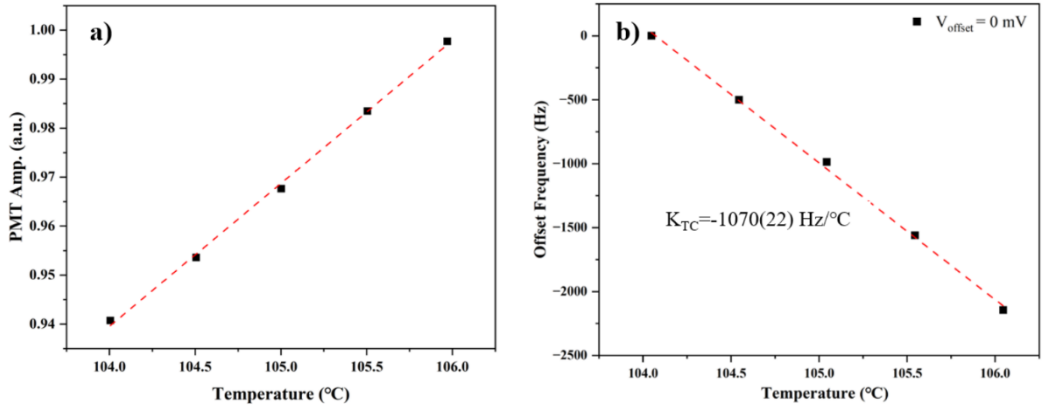

FIG. 6. (a) Experimetally measured 420 nm fluoresonce amplitude depends on the vapor-cell temperature. The red dash line is the linear fitting. (b) Experimentally measured cell-tempearature sensitivity (on resonance locking, $V_{offset}$ = 0 V), the fitting slope is -1070 (20) Hz/°C @ 778 nm.

Figure 7 presents the temperature frequency shift curves under various offset voltages, showing distinct behaviors.

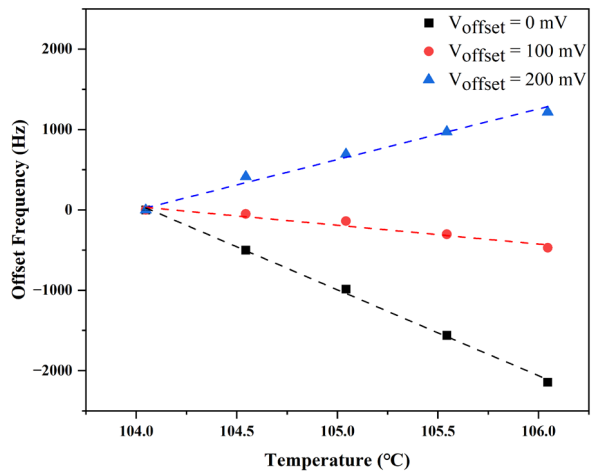

FIG. 7 (a) Measured offset frequency depends on the vapor-cell temperature on-resonance locking ($V_{offset}$ = 0 mV, black), with compensation ($V_{offset}$ = 100 mV, red) and overcompensation ($V_{offset}$ = 200 mV, blue). Offset frequency data operated under 104 °C are set to "Zero" offset frequency.

With on-resonance locking ($V_{offset}$ = 0 mV), there is a -1068(20) Hz/°C shift coefficient (black fitted line). Meanwhile, with compensation ($V_{offset}$ = 100 mV), a minimal shift sensitivity of about -233(28) Hz/°C (red dash line) is observed, demonstrating effective temperature shift mitigation via offset locking. Interestingly, an offset of 200 mV leads to overcompensation, flipping the shift coefficient to +1000(10) Hz/°C (blue dash line). This identifies $V_{offset}$ ≈100 mV as a good point for mitigating temperature sensitivity, essentially neutralizing its impact on frequency shifts.

To assess real-time shift compensation, we have monitored the output frequency of the 2hv-ROFS as it responds to a controlled cell temperature drift, gradually increasing at a rate of about 6.7 mK per second from 104 °C to 106 °C over a period of 3000 seconds (refer to Fig. 7a). This temperature drift was tailored to align with the thermal response time of the vapor cell, guaranteeing 'quasi-static' frequency shift compensation. The optical comb was referenced to an H-maser, instead of the super-cavity-stabilized laser, for the long-term frequency drift measurement. Figures 7b and 7c provide additional details regarding the system's output frequency and stability in response to cell temperature drift, with and without compensation. Additionally, the comparative analysis showcases the typical output frequency performance under a constant cell temperature condition of 104 °C. Under the on-resonance locking condition ($V_{offset}$ = 0 V), the output frequency exhibits a total of approximately 2 kHz drift in response to temperature drifts. Meanwhile, under the optimal offset point ($V_{offset}$ = 110 mV), the drift is substantially minimized to approximately 300 Hz. The corresponding frequency stability also shows similar results. The compensation effect begins to appear after an average sampling time of 10 seconds. Beyond an average sampling time of 100 seconds, when the frequency instability caused by temperature drift becomes dominant, the stability under compensation improves by approximately seven times compared to that without compensation. We believe that if $V_{offset}$ can be further finely optimized, it could ultimately achieve "zero" temperature sensitivity of the vapor cell.

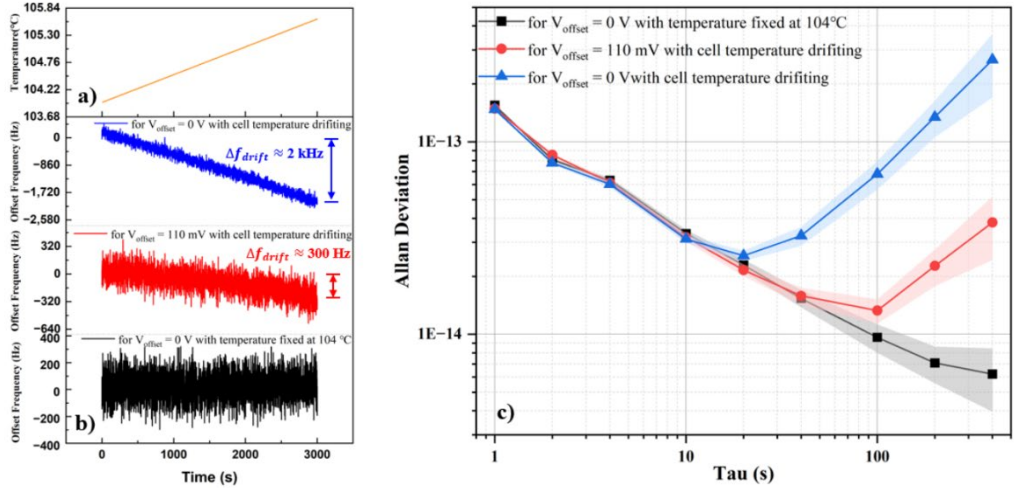

FIG. 8. (a) Intentially controlled real-time cell temperature drift. (b) Real-time output freuqency operated for $V_{offset}$ = 0 V (blue) and $V_{offset}$ = 110 mV (red) with cell temperature drfiting, and for comparision the real-time output freuqency for $V_{offset}$ = 0 V with temperarute fixed at 104°C (black) is shown. (c) the corresponding stability performances for the three conditions in (b).

## V. DISCUSSION AND CONCLUSION

Compared to the on-resonance locking scheme, offset locking introduces an additional frequency shift determined by both the offset voltage ($V_{offset}$) and the amplitude of atomic resonance. However, in many applications of compact atomic clock technology, long-term stability and frequency traceability are prioritized over frequency accuracy. Concerning the instability of this additional shift, the temperature sensitivity of the atomic resonance's amplitude may deserve more attention than the instability of $V_{offset}$. The amplitude is influenced by the electronic gains, including the efficiency of the PMT, the gain of the Lock-in, and the preamplifier. Any changes in gain due to environmental temperature variations will degrade the clock's long-term stability. We are currently researching the stability of these parameters on the clock platform. Regarding shift traceability, since $V_{offset}$ can be maintained with high precision by the electronics, its impact on frequency repeatability is negligible. For instance, when suppressing the light shift, the additional offset frequency shift was approximately 1 kHz ($V_{offset} \approx 30$ mW). We conducted on-off locking cycles (while keeping the vapor cell heated) and tested the offset-locking absolute frequency over several days. The measured retrace was ~5E-13, comparable to the level operated under on-resonance conditions [36].

It is also important to note the polarity with light shifts and vapor-cell temperature shifts. If these two types of frequency shifts have the same polarity, then offset locking can simultaneously reduce the frequency shift coefficients of both effects; if the polarities are opposite, it might exacerbate the coefficient of the other frequency shift. In the two-photon frequency standard, the polarities of light frequency shifts and temperature frequency shifts are the same, so offset locking positively affects the suppression of sensitivity to both types of frequency shifts.

Finally, to achieve the perfect compensation effect, we should ensure that the offset locking shift is significantly smaller than the linewidth of the atomic resonance and that variations in environmental parameters are not too large. Otherwise, the presence of nonlinear frequency shifts during the compensation process will impact the effectiveness of the compensation.

In this paper, we propose a ROL method to compensate for both light shifts and vapor-cell temperature shifts. We have successfully demonstrated the suppression of these two shift sensitivities on a 778 nm 2hv-ROFS. This approach is straightforward to implement and does not increase the system's SWaP and complexity. It can effectively counteract frequency shifts caused by factors including vapor-cell temperature, the intensity of pumping or probing lasers, and microwave power, as long as the atomic resonance depends on the environmental factor. This technique is particularly well-suited for compact atomic clocks that prioritize long-term frequency stability over high accuracy, such as Rb clocks, chip-scale CPT clocks, and vapor-cell optical clocks based on saturated absorption, two-photon, and modulation transfer spectroscopy techniques.

## ACKNOWLEDGMENTS

We acknowledge financial support by Chinese Academy of Sciencies and thank Zach Newman for helpful discussions.

## REFERENCES

[1] P. Forman, The Quantum Physics of Atomic Frequency Standards, Chap.1, pp. 69-82.

[2] W. J. Riley, A History of the Rubidium Frequency Standard, IEEE UFFC-S History, http://ieee-uffc.org/aboutus/history/a-history-of-the-rubidium-frequency-standard.pdf, Dec. 2019.

[3] Forman P, The Quantum Physics of Atomic Frequency Standards, Chap.2, pp.127-135.

[4] S. Knappe, V. Shah, P. D. Schwindt, L. Hollberg, J. Kitching, L. A. Liew, and J. Moreland, A microfabricated atomic clock. Applied Physics Letters, 85, 9(2004).

[5] L. A. Mallette, P. Rochat and J. White, Historical Review of Atomic Frequency Standards Used in Satellite Based Navigation Systems, in Proceedings of the 63rd Annual Meeting of The Institute of Navigation, ION(2007), pp. 40-48.

[6] K. F. Hasan, C. Wang, Y. Feng and Y. C. Tian, Time synchronization in vehicular ad-hoc networks: A survey on theory and practice. Vehicular communications, 14(2018).

[7] R. Morello, S. C. Mukhopadhyay, Z. Liu, D. Slomovitz and S. R. Samantaray, Advances on sensing technologies for smart cities and power grids: A review. IEEE Sensors Journal, 17, 23(2017).

[8] T. Bandi, C. Affolderbach, C. Stefanucci, F. Merli, A. K. Skrivervik and G. Mileti, Compact high-performance continuous-wave double-resonance rubidium standard with $1.4\times10^{-13}\tau^{-1/2}$ stability. IEEE transactions on ultrasonics, ferroelectrics, and frequency control, 61, 11(2014).

[9] M. Gozzelino, S. Micalizio, C. E. Calosso, J. Belfi, A. Sapia, M. Gioia and F. Levi, Realization of a pulsed optically pumped Rb clock with a frequency stability below 10-15. Scientific Reports, 13, 1(2023).

[10] Q. Hao, S. Yang, J. Ruan, P. Yun and S. Zhang, Integrated pulsed optically pumped Rb atomic clock with frequency stability of 10−15. Physical Review Applied, 21, 2(2024).

[11] M. Abdel Hafiz, G. Coget, P. Yun, S. Guérandel, E. de Clercq and R. Boudot, A high-performance Raman-Ramsey Cs vapor cell atomic clock. Journal of Applied Physics, 121, 10(2017).

[12] K. W. Martin, G. Phelps, N. D. Lemke, M. S. Bigelow, B. Stuhl, M. Wojcik, M. Holt, I. Coddington, M. W. Bishop, and J. H. Burke, Compact optical atomic clock based on a two-photon transition in rubidium. Physical Review Applied, 9, 1(2018).

[13] J. D. Roslund et al., Optical clocks at sea. Nature, 628, 8009(2024).

[14] Z. L. Newman, V. Maurice, C. Fredrick, T. Fortier, H. Leopardi, L. Hollberg, S. A. Diddams, J. Kitching and M. T. Hummon, High-performance,

compact optical standard. Optics Letters, 46, 18(2021).

[15] V. Formichella, J. Camparo, I. Sesia, G. Signorile, L. Galleani, M. Huang and P. Tavella, The ac Stark shift and space-borne rubidium atomic clocks. Journal of Applied Physics, 120, 19(2016).

[16] D. Miletic, C. Affolderbach, G. Mileti, M. Hasegawa and C. Gorecki, Light shift in CPT based cs miniature atomic clocks. In 2011 Joint Conference of the IEEE International Frequency Control and the European Frequency and Time Forum (FCS) Proceedings, pp. 1-3.

[17] C. Affolderbach, C. Andreeva, S. Cartaleva, T. Karaulanov, G. Mileti and D. Slavov, Light-shift suppression in laser optically pumped vapour-cell atomic frequency standards. Applied Physics B, 80(2005).

[18] C. Perrella, P. S. Light, J. D. Anstie, F. N. Baynes, R. T. White and A. N. Luiten, Dichroic two-photon rubidium frequency standard. Physical Review Applied, 12, 5(2019).

[19] Q. Hao, W. Xue, F. Xu, K. Wang, P. Yun and S. Zhang, Efforts towards a low-temperature-sensitive physics package for vapor cell atomic clocks. Satellite Navigation, 1(2020).

[20] C. E. Calosso, A. Godone, F. Levi and S. Micalizio, Enhanced temperature sensitivity in vapor-cell frequency standards. IEEE transactions on ultrasonics, ferroelectrics, and frequency control, 59, 12(2012).

[21] T. N. Nguyen and T. R. Schibli, Temperature-shift-suppression scheme for two-photon two-color rubidium vapor clocks. Physical Review A, 106, 5(2022).

[22] O. Kozlova, J. M. Danet, S. Guérandel and E. de Clercq, Temperature dependence of a Cs vapor cell clock: Pressure shift, signal amplitude, light shift. In 2011 Joint Conference of the IEEE International Frequency Control and the European Frequency and Time Forum (FCS) Proceedings, pp. 1-5.

[23] G. H. MEI et al., Characteristics of the space-borne rubidium atomic clocks for the BeiDou III navigation satellite system. SCIENTIA SINICA Physica, Mechanica & Astronomica, 51, 1(2021).

[24] J. Vanier and C. Audoin, The Quantum Physics of Atomic Frequency Standards, Vol.1, Chap.7, pp. 1310-1380.

[25] D. Miletic, Light-shift and temperature-shift studies in atomic clocks based on coherent population trapping (Doctoral dissertation), Chap.4, pp. 97-102.

[26] V. Gerginov and K. Beloy, Two-photon optical frequency reference with active ac Stark shift cancellation. Physical Review Applied, 10, 1(2018).

[27] E. A. Tsygankov, D. S. Chuchelov, M. I. Vaskovskaya, V. V. Vassiliev, S. A. Zibrov and V. L. Velichansky, A nonlinear frequency shift caused by asymmetry of the coherent population trapping resonance: a generalization. arXiv preprint arXiv: 2311.17229. 2023 Nov 28.

[28] V. I. Yudin, M. Y. Basalaev, A. V. Taichenachev, O. N. Prudnikov, D. A. Radnatarov, S. M. Kobtsev, S. M. Ignatovich and M. N. Skvortsov, Frequency shift caused by the line-shape asymmetry of the resonance of coherent population trapping. Physical Review A, 108, 1(2023).

[29] D. V. Brazhnikov, S. M. Ignatovich and M. N. Skvortsov, Light Shift Suppression in Coherent-Population-Trapping Atomic Clocks in the Field of Two Circularly Polarized Light Beams. arXiv preprint arXiv:2311.00461. 2023 Nov 1.

[30] V. I. Yudin, M. Y. Basalaev, A. V. Taichenachev, O. N. Prudnikov, D. A. Radnatarov, S. M. Kobtsev, S. M. Ignatovich and M. N. Skvortsov, Frequency shift caused by the line-shape asymmetry of the resonance of coherent population trapping. Physical Review A, 108, 1(2023).

[31] V. I. Yudin et al., General methods for suppressing the light shift in atomic clocks using power modulation. Physical Review Applied, 14, 2(2020).

[32] M. A. Hafiz, R. Vicarini, N. Passilly, C. E. Calosso, V. Maurice, J. W. Pollock, A. V. Taichenachev, V. I. Yudin, J. Kitching and R. Boudot, Protocol for light-shift compensation in a continuous-wave

microcell atomic clock. Physical Review Applied, 14, 3(2020).

[33] M. Abdel Hafiz, C. Carlé, N. Passilly, J. M. Danet, C. E. Calosso and R. Boudot, Light-shift mitigation in a microcell-based atomic clock with symmetric auto-balanced Ramsey spectroscopy. Applied Physics Letters, 120, 6(2022).

[34] D. Li, K. Liu, P. Wang and S. Kang, Dual-interrogation method for suppressing light shift in Rb 778 nm two-photon transition optical frequency standard. Optics Express, 32, 2(2024).

[35] G. Grynberg and B. Cagnac, Doppler-free multiphotonic spectroscopy. Reports on Progress in Physics, 40, 7(1977).

[36] M. F. Qu, D. Li, C. H. Li, K. Q. Liu, W. H. Zhu, Y. Wei, P. F. Wang and S. B. Kang, Towards a compact soliton microcomb fully referenced on atomic reference. arxiv preprint arxiv:2310.08957 (2023).